
\def\leaderdot{\leaders\hbox to 1 em {\hss.\hss}\hfill}

\def\dlowlim#1#2{\mathop{\rm #1}\limits_{#2}}

\dimen0= \parindent
\dimen1= \hsize \advance\dimen1 by -\dimen0

\dimen2=\baselineskip
\def\skiplines#1 { \dimen3=\dimen2 \multiply\dimen3 by #1 \vskip \dimen3}
\def\fullline{\hbox to \fullhsize}

\def\numpage{\baselineskip=24pt\fullline{\the\footline}}

\def\mathcedilla{\vtop{\hbox{c}{\kern0pt\nointerlineskip}
	         {\hbox{$\mkern-2mu \mathchar"0018\mkern-2mu$}}}}

\mathchardef\gq="0060
\mathchardef\dq="0027
\mathchardef\ssmath="19
\mathchardef\aemath="1A
\mathchardef\oemath="1B
\mathchardef\omath="1C
\mathchardef\AEmath="1D
\mathchardef\OEmath="1E
\mathchardef\Omath="1F
\mathchardef\imath="10
\mathchardef\fmath="0166
\mathchardef\gmath="0167
\mathchardef\vmath="0176


\def\colleft{\strut\kern.3em}
\def\colright{\kern0pt}

\def\figureh{\hbox to}

\def\m@th{\mathsurround=0pt}
\newif\ifdtpt
\def\displ@y{\openup1\jot\m@th
    \everycr{\noalign{\ifdtpt\dt@pfalse
    \vskip-\lineskiplimit \vskip\normallineskiplimit
    \else \penalty\interdisplaylinepenalty \fi}}}
\def\eqalignl#1{\,\vcenter{\openup1\jot\m@th
                \ialign{\strut$\displaystyle{##}$\hfil&
                              $\displaystyle{{}##}$\hfil&
                              $\displaystyle{{}##}$\hfil&
                              $\displaystyle{{}##}$\hfil&
                              $\displaystyle{{}##}$\hfil\crcr#1\crcr}}\,}
\def\eqalignnol#1{\displ@y\tabskip\centering \halign to \displaywidth{
                  $\displaystyle{##}$\hfil\tabskip=0pt &
                  $\displaystyle{{}##}$\hfil\tabskip=0pt &
                  $\displaystyle{{}##}$\hfil\tabskip=0pt &
                  $\displaystyle{{}##}$\hfil\tabskip=0pt &
                  $\displaystyle{{}##}$\hfil\tabskip\centering &
                  \llap{$##$}\tabskip=0pt \crcr#1\crcr}}
\def\leqalignnol#1{\displ@y\tabskip\centering \halign to \displaywidth{
                   $\displaystyle{##}$\hfil\tabskip=0pt &
                   $\displaystyle{{}##}$\hfil\tabskip=0pt &
                   $\displaystyle{{}##}$\hfil\tabskip=0pt &
                   $\displaystyle{{}##}$\hfil\tabskip=0pt &
                   $\displaystyle{{}##}$\hfil\tabskip\centering &
                   \kern-\displaywidth\rlap{$##$}\tabskip=\displaywidth
                   \crcr#1\crcr}}
\def\eqalignc#1{\,\vcenter{\openup1\jot\m@th
                \ialign{\strut\hfil$\displaystyle{##}$\hfil&
                              \hfil$\displaystyle{{}##}$\hfil&
                              \hfil$\displaystyle{{}##}$\hfil&
                              \hfil$\displaystyle{{}##}$\hfil&
                              \hfil$\displaystyle{{}##}$\hfil\crcr#1\crcr}}\,}
\def\eqalignnoc#1{\displ@y\tabskip\centering \halign to \displaywidth{
                  \hfil$\displaystyle{##}$\hfil\tabskip=0pt &
                  \hfil$\displaystyle{{}##}$\hfil\tabskip=0pt &
                  \hfil$\displaystyle{{}##}$\hfil\tabskip=0pt &
                  \hfil$\displaystyle{{}##}$\hfil\tabskip=0pt &
                  \hfil$\displaystyle{{}##}$\hfil\tabskip\centering &
                  \llap{$##$}\tabskip=0pt \crcr#1\crcr}}
\def\leqalignnoc#1{\displ@y\tabskip\centering \halign to \displaywidth{
                  \hfil$\displaystyle{##}$\hfil\tabskip=0pt &
                  \hfil$\displaystyle{{}##}$\hfil\tabskip=0pt &
                  \hfil$\displaystyle{{}##}$\hfil\tabskip=0pt &
                  \hfil$\displaystyle{{}##}$\hfil\tabskip=0pt &
                  \hfil$\displaystyle{{}##}$\hfil\tabskip\centering &
                  \kern-\displaywidth\rlap{$##$}\tabskip=\displaywidth
                  \crcr#1\crcr}}
\def\dasharrowfill{$\mathsurround=0pt \mathord- \mkern-6mu
    \cleaders\hbox{$\mkern-2mu\mathord-\mkern-2mu$}\hfill
    \mkern-6mu \mathord-$}



\def\charlvmidlw#1#2{\,\vtop{\ialign{##\crcr
      #1\crcr\noalign{\kern1pt\nointerlineskip}
      $\hfil#2\hfil$\crcr}}\,}
\def\charlvlowlw#1#2{\,\vtop{\ialign{##\crcr
      $\hfil#1\hfil$\crcr\noalign{\kern1pt\nointerlineskip}
      #2\crcr}}\,}
\def\charlvmidup#1#2{\,\vbox{\ialign{##\crcr
      $\hfil#1\hfil$\crcr\noalign{\kern1pt\nointerlineskip}
      #2\crcr}}\,}
\def\charlvupup#1#2{\,\vbox{\ialign{##\crcr
      #1\crcr\noalign{\kern1pt\nointerlineskip}
      $\hfil#2\hfil$\crcr}}\,}

\def\vspce{\kern4pt} \def\hspce{\kern4pt}    

\def\emptybox{\vbox{\kern.7ex\hbox{\kern.5em}\kern.7ex}}
 \font\sevmi  = cmmi7              
    \skewchar\sevmi ='177          
 \font\fivmi  = cmmi5              
    \skewchar\fivmi ='177          
\font\tenmib=cmmib10
\newfam\bfmitfam

\textfont\bfmitfam=\tenmib
\scriptfont\bfmitfam=\sevmi
\scriptscriptfont\bfmitfam=\fivmi


\def\twodot{.\kern-0.1em.}

\def\paral{\mathrel{/\kern-.25em/}}
\def\grlo{\mathrel{\hbox{\lower.2ex\hbox{\rlap{$>$}\raise1ex\hbox{$<$}}}}}
\def\logr{\mathrel{\hbox{\lower.2ex\hbox{\rlap{$<$}\raise1ex\hbox{$>$}}}}}
\def\greq{\mathrel{\hbox{\lower1ex\hbox{\rlap{$=$}\raise1.2ex\hbox{$>$}}}}}
\def\loeq{\mathrel{\hbox{\lower1ex\hbox{\rlap{$=$}\raise1.2ex\hbox{$<$}}}}}
\def\grsim{\mathrel{\hbox{\lower1ex\hbox{\rlap{$\sim$}\raise1ex\hbox{$>$}}}}}
\def\losim{\mathrel{\hbox{\lower1ex\hbox{\rlap{$\sim$}\raise1ex\hbox{$<$}}}}}
\font\ninerm=cmr9
\def\uniset{\rlap{\ninerm 1}\kern.15em 1}

\def\emptysq{\mathbin{\vbox{\hrule\hbox{\vrule height1ex \kern.5em
                            \vrule height1ex}\hrule}}}
\def\emptyrect{\mathbin{\vbox{\hrule\hbox{\vrule height1ex \kern1em
                              \vrule height1ex}\hrule}}}
\def\rightonleftarrow{\mathrel{\hbox{\raise.5ex\hbox{$\rightarrow$}\ignorespaces
                                   \lower.5ex\hbox{\llap{$\leftarrow$}}}}}
\def\leftonrightarrow{\mathrel{\hbox{\raise.5ex\hbox{$\leftarrow$}\ignorespaces
                                   \lower.5ex\hbox{\llap{$\rightarrow$}}}}}

\def\bkB{{\rm I\kern-.17em B}}
\def\bkC{{\rm \kern.24em
            \vrule width.05em height1.4ex depth-.05ex
            \kern-.26em C}}
\def\bkD{{\rm I\kern-.17em D}}
\def\bkE{{\rm I\kern-.17em E}}
\def\bkF{{\rm I\kern-.17em F}}
\def\bkG{{\rm \kern.24em
            \vrule width.05em height1.4ex depth-.05ex
            \kern-.26em G}}
\def\bkH{{\rm I\kern-.22em H}}
\def\bkI{{\rm I\kern-.22em I}}
\def\bkJ{{\rm \kern.19em
            \vrule width.02em height1.5ex depth0ex
            \kern-.20em J}}
\def\bkK{{\rm I\kern-.22em K}}
\def\bkL{{\rm I\kern-.17em L}}
\def\bkM{{\rm I\kern-.22em M}}
\def\bkN{{\rm I\kern-.20em N}}
\def\bkO{{\rm \kern.24em
            \vrule width.05em height1.4ex depth-.05ex
            \kern-.26em O}}
\def\bkP{{\rm I\kern-.17em P}}
\def\bkQ{{\rm \kern.24em
            \vrule width.05em height1.4ex depth-.05ex
            \kern-.26em Q}}
\def\bkR{{\rm I\kern-.17em R}}
\def\bkT{{\rm \kern.24em
            \vrule width.02em height1.5ex depth 0ex
            \kern-.27em T}}
\def\bkU{{\rm \kern.30em
            \vrule width.02em height1.47ex depth-.05ex
            \kern-.32em U}}
\def\bkZ{{\rm Z\kern-.32em Z}}



%
\def\midfig#1x#2:#3#4{\midinsert
$$\vbox to #2{\hbox to #1{\special{ps:#3}\hfill}\vfill}$$\par
#4\par\endinsert}
\def\topfig#1x#2:#3#4{\topinsert
$$\vbox to #2{\hbox to #1{\special{ps:#3}\hfill}\vfill}$$\par
#4\par\endinsert}
\def\infig#1x#2:#3{
$$\vbox to #2{\hbox to #1{\special{ps:#3}\hfill}\vfill}$$}
\def\textfig#1x#2:#3{
$\vbox to #2{\hbox to #1{\special{ps:#3}\hfill}\vfill}$}
\magnification=1200
\baselineskip=17pt
\ \bigskip
\centerline{\bf Phase diagram of a semiflexible polymer chain in a $\theta$
solvent: application to protein folding}
\par
\vskip 4mm
\centerline{by}
\vskip 3mm \centerline{ S.Doniach \footnote
{$^+$}{usually at: Dept. Applied Physics, Stanford University,
Stanford CA 94305}, T. Garel, and H.Orland\footnote {$ ^\ast
$}{Also at Groupe de Physique Statistique, Universit\'e de
Cergy-Pontoise, 95806 Cergy-Pontoise Cedex, France}}
\centerline{{\sl Service de Physique Th\'eorique\/}\footnote {$ ^{\dagger}
$}{Laboratoire de la Direction des Sciences de la Mati\`ere du Commissariat
\`a l'Energie Atomique}}
\centerline{{\sl CE-Saclay, 91191 Gif-sur-Yvette Cedex, France\/}}
\par
\vskip 50mm
\noindent{\lq\lq Submitted for publication to \sl {J. Chem.Phys.}\rq\rq}
\noindent \hfill{Saclay, SPhT/95-119} \par
\vskip 30mm
Short title: Phase diagram of a protein chain.\par
\par
\vfill\eject
{\bf ABSTRACT}
\par
We consider a lattice model of a semiflexible homopolymer chain in a bad
solvent. Beside the temperature $T$, this model is described by (i) a
curvature energy $\varepsilon_h$, representing the stiffness of the
chain (ii) a nearest-neighbour attractive energy $\varepsilon_v$,
representing the solvent (iii) the monomer density $\rho={N \over
\Omega}$, where $N$ and $\Omega$ denote respectively the number of
monomers and the number of lattice sites.  This model is a simplified
view of the protein folding problem, which encompasses the geometrical
competition between secondary structures (the curvature
term modelling helix formation)
and the global compactness (modeled here by the attractive energy),
but contains no side chain information.
By  allowing the monomer density $\rho$ to depart from unity one has
made a first (albeit naive) step to include the role of the water.
In previous analytical studies we considered only the (fully compact)
case $\rho=1$,
and found a first order freezing transition towards a crystalline ground
state (also called the native state in the protein literature).\par
 In this paper, we extend this calculation to the description of both compact
and non-compact phases. The analysis is done first at a mean-field level.
We then
find that the transition from the high temperature swollen coil state to the
crystalline ground state is a two-step process for which:(i) there is first a
$\theta$ collapse transition towards a compact ``liquid'' globule, and  (ii) at
low temperature, this ``liquid'' globule undergoes a discontinuous
freezing transition.
The mean-field value of the $\theta$ collapse temperature is found to
be independent of the curvature energy $\varepsilon_h$.
This mean-field analysis is improved by a variational bound, which
confirms the independence of the $\theta$ collapse temperature with
respect to $\varepsilon_h$. This result is confirmed by a Monte Carlo
simulation, although with a much lower value of the $\theta$ temperature.
This lowering of the collapse transition allows the possibility (for
large $\varepsilon_h$)
of a
direct first order freezing transition, from a swollen coil to the
crystalline ground state. For small values of $\varepsilon_h$, the
mean-field two-step mechanism remains valid.

In the protein folding problem, the ``liquid''
compact phase is likely to be related to the ``molten globule'' phase.
The properties of this model system thus suggest that, even though
side chain disordering is not taken into account, disordering of the
backbone of a protein may  still be a sufficient mechanism to drive the system
from the native state into the  molten globule state.
\vfill\eject

\noindent {\bf 1.\nobreak\ INTRODUCTION}
\par
The thermodynamic equilibrium states of small globular proteins in solution
depend on solvent conditions, including:  temperature, pH, salt concentration
and denaturant concentration.
\par
Roughly speaking, there are two ways to go from a fully unfolded swollen
coil state to the native state:  through a single first order folding
transition, or through an intermediate collapsed phase [1].  The nature of
the transition from an unfolded state to this phase, which may be
generically called a molten globule [2,3], depends in detail on the specific
protein and solvent conditions, but often has the character of a
continuous $\theta$-collapse.  The subsequent transition to the native state is
generally first order in character.
Note that, in a finite system such as a protein, the use of the
expression ``phase transition" should be interpreted as ``strongly
cooperative phenomenon."
\par
In order to elucidate the balance of enthalpic
and entropic terms leading to this phase diagram we have previously
represented the protein folding process using a highly simplified
homopolymer model, originally due to Flory [4].  In this model the tendency
to form local order which may have 1-or 2-dimensional character ($\alpha$
helices or $\beta$ sheets) is complemented by a tendency to global compactness
representing the effect of hydrophobic forces. The properties of this
model were previously studied numerically by Kolinsky et al [5].
We have previously given
an analytical mean field treatment of the model [6,7] which was restricted to
the fully compact phases and yielded a first order freezing transition
towards a native-like structure.
\par
In this paper we extend this model by (i) allowing for the presence of
vacancies (ii) assigning a finite attractive energy $\varepsilon_v$ between
(nonconsecutive) neighbouring monomers.  This generalization enables us
to study the full phase diagram of the model by the method of mean field
theory, including the high temperature disordered coil state. From a
more (bio)chemical point of view, points (i) and (ii) above are a
first step towards a proper treatment of the solvent properties.
\par
We report elsewhere the results of numerical studies which yield information
about the structural character of the partially folded states of this
model [8].
\par
Besides its application for proteins, this model is clearly of interest
to describe the melting or crystallization of semi-flexible polymers [4]:
in this case, intermediate compact phases are likely to possess
orientational order [9,10].
\par
The organization of this paper is as follows.
In section 2, we recall for completeness our previous study of fully
compact phases, through the formalism of Hamiltonian paths.  The method
is applied to the case of 1-dimensional structures which may be thought
of as helix-like (sheets may be treated in a similar way).  The extended
model, also restricted to the case of helix-like structures, is presented
at a mean-field level in section 3, together with its resulting phase diagram.
In section 4, we derive an upper bound for the free energy, which
strongly suggests that the collapse temperature is indeed independent
of the curvature energy. Using previous determinations of the $\theta
$ temperature, we obtain a modified phase diagram, which is then
checked numerically by Monte Carlo simulations.
Connections
with other approaches are considered in the conclusion.
\vskip 3mm
\noindent{\bf 2. SECONDARY STRUCTURES IN A COMPACT PHASE} \par
\smallskip
 In this section, we shall describe how one can study, in terms
of a very simplified and schematic homopolymer model the thermodynamics
of secondary structures in a compact phase. Before doing so, let us
reconsider the problem of the collapsed chain on a lattice, initially
without introducing any tendency to secondary structure formation (i.e.
for zero stiffness).
\smallskip
{\bf 2.1} {\bf Hamiltonian paths and the compact phase} \par
\smallskip
We consider a
$ d $-dimensional hypercubic lattice of $ N=L^d $ sites,
and its associated Hamiltonian paths representing the conformations of a
polymer with $N-1$ links. We
recall that a Hamiltonian path (HP) is a path which
visits all sites of the lattice
once and only once . In this approach, to compute the
entropy of the collapsed but disordered phase means to evaluate the number
of Hamiltonian paths on the lattice. For
simplicity, we consider closed paths, but, as is well known in polymer
theory [11], (i.e. in the large $N$ limit), boundary conditions play only a
subdominant role in terms of the free energy.
\par
Thus we define the partition function $\cal Z$ by:
$$ {\cal Z }= \sum_{\{HP\}}{\rm \bf 1} \eqno (2.1) $$
Note that ${\cal Z }$ is nothing but the total number of HP.
One may rewrite (2.1) as:
$$ {\cal Z }= \lim_{n \to 0} {1 \over n} { \int
 \prod_{\{ \vec r \}} {\rm d}\vec  \varphi  \left(\vec  r \right)
{\rm e}^{-A_G} \prod_{\{ \vec r \} } { 1 \over 2} \vec
\varphi^{ \ 2} \left(\vec  r \right)
 \over {\int \prod_
{\{ \vec r \} } {\rm d}\vec \varphi \left(\vec  r \right) {\rm e}^{-A_G}}}
\eqno (2.2.a) $$
where
$$ A_G = {1 \over 2} \sum_{ rr^{\prime}}\vec
\varphi \left(\vec  r \right) \left(\Delta_{\vec  r\vec  r^{\
\prime}} \right)^{-1}\vec  \varphi\left(\vec  r^{\ \prime} \right)
\eqno (2.2.b) $$
In eq.(2.2), $\varphi(\vec r)$ is a $n-$component field attached to each
point $\vec r$ of the lattice, and the operator $\Delta_{\vec  r\vec  r^{\
\prime}}$ is $1$ if $\vec r$ and $\vec r^{\prime}$ are
nearest-neighbour, and $0$ otherwise.
The denominator is a simple Gaussian integral, equal to $\exp(-{n\over 2}
\log \det \Delta_{\vec  r\vec  r^{\prime}} )$. It is equal to one in
the limit $ n \to 0$, and will be omitted henceforth.
\par
To prove the identity between
(2.2) and (2.1), namely that (2.2) indeed counts all HP, we use
Wick's theorem. At each site $\vec
r$, one
attaches a factor ${ 1 \over 2} \vec \varphi^{ \ 2} \left(\vec  r
\right) $, which can be viewed graphically as one line coming into
, and one line coming out of site $\vec r$. The elementary contraction
(see eq. (2.2.b)) reads:
$$ \charlvupup{ \dasharrowfill}{ \varphi_\alpha \left(\vec  r
\right)\cdot \varphi_\beta \left(\vec  r^{\ \prime} \right)} =
\delta_{ \alpha \beta} \Delta_{\vec  r\vec  r^{\
\prime}}
\eqno(2.3)
$$
where $\alpha$ and $\beta$ denote any of the $n-$components of the
field ${\vec \varphi}(\vec r)$. Since eq.(2.3) conserves the index
$\alpha$, each connected part of the path yields a factor $n$ through
the summation over $\alpha$. To deal with the one chain problem (or
with a single connected path), one must therefore extract the term
proportionnal to $n$, hence the factor $\lim_{n \to
0} {1 \over n}$ in eq.(2.2.a). By doing so [9], paths with more than one
connected part, which  have higher powers of $n$, are cancelled, which
proves the identity of (2.1) and (2.2).
\par
The exact evaluation of (2.2) is out of reach, and we restrict ourselves to
a simple approximation, namely the Saddle-Point Method (SPM).
Minimizing the exponent of the numerator of (2.2.a)  with respect to
${\vec \varphi}(\vec r)$ yields:
$$
\sum_{\vec r \prime} \Delta_{\vec  r\vec  r^{
\ \prime}} ^{-1}\vec  \varphi\left(\vec  r^{\ \prime} \right) =
{2 \vec \varphi(\vec r) \over {\vec \varphi(\vec r)}^2}
\eqno (2.4)
$$
If we assume periodic boundary conditions on the lattice, it is
legitimate to look for a homogeneous solution to (2.4).
We further break the $ O(n) $ symmetry by choosing $ \vec
\varphi $ in a given
direction, say 1, and a take the solution as:
$$
\vec \varphi({\vec r}) = ( \varphi, 0, \ldots)
\eqno(2.5.a)
$$
where $\varphi$ is a constant, independent of space. Using (2.4) we
get $ \varphi^2 = 2 q$, where $q=2d$ is the coordination number of the
lattice. In the SPM, the partition function (2.4) reads:
$$
Z= \left( {q \over {\rm e}} \right)^N
\eqno(2.5.b)
$$
where ${\rm e}=\exp(1)=2.718\ldots $. This result [12] is in excellent
agreement
with numerical simulations and exact results, on various lattices,
even in low dimensions. This is due to the fact that fluctuations
around the SPM are indeed small [2,12].
\par
Equation (2.5.b) shows that in the limit of zero stiffness
the compact phase has a sizeable conformational entropy $ S/N = \log(q/{\rm
e})$
, and thus no
definite three-dimensional shape: a polymer globule (compact phase)
is a statistical mixture of an exponentially large number of
conformations. (We remind the reader that these results are strongly
dependent on the boundary conditions, assumed here to be periodic).
To sum up, we have modeled the collapsed phase of a single polymer
chain below the
$\theta$ point by lattice Hamiltonian paths. In this approach, one
looses track of the relation between the spatial degrees of freedom
and the curvilinear abscissa along the chain.
\par
\smallskip
{\bf 2.2)} {\bf Secondary structure in the compact phase} \par
\smallskip
Following Flory
[4], we
may think of each link of the Hamiltonian path as representing a helical
turn. Since hydrogen-bonds have a tendency to favor long helices,
that is to align the links of our model, we attribute an energy
penalty $ \varepsilon_h $ to the breaking of an helix, that is whenever
the Hamiltonian path makes a turn (corner). This model of a homopolymer with
finite stiffness has attracted a lot of attention in the theory of
polymer melting [6,9,10].
The partition function of
the system, at inverse temperature $ \beta  = {1 \over T}, $ reads
\smallskip
$$ {\cal Z }= \sum^{ }_{\{HP\}} {\rm e}^{-\beta \varepsilon_h N_c\{{\cal
H}\}} \eqno (2.6) $$
\noindent where ${\{HP\}} $ denotes the ensemble of all Hamiltonian
paths, and $ N_c\{{\cal H}\} $
denotes the number of corners present in path $ {\cal H}. $ Following
the previous section,
we introduce on each lattice site $ \vec  r $ and
for each direction $ \alpha =1,...,d\  $ an $ n $-component real field $ \vec
\varphi_ \alpha \left(\vec  r \right). $ The
partition function $ Z $ can be rewritten:
$$ {\cal Z} = \dlowlim{ {\rm lim}}{n \longrightarrow 0} {1 \over n} \int^{ }_{
}
\prod^{ }_{\vec  r} \prod^ d_{\alpha =1} {\rm d}\vec  \varphi_ \alpha
\left(\vec  r \right) {\rm e}^{-A_G} \prod^{ }_{\vec  r} \left[{1 \over 2}
\sum^ d_{\alpha =1}\vec  \varphi^ 2_\alpha \left(\vec  r \right)+ {\rm
e}^{-\beta \varepsilon_h} \sum^{ }_{ 1\leq \alpha <\gamma \leq d}\vec  \varphi_
\alpha \left(\vec  r \right)\cdot\vec  \varphi_ \gamma \left(\vec  r \right)
\right] \eqno (2.7.a) $$
with
$$ A_G = {1 \over 2} \sum^ d_{\alpha =1} \sum^{ }_{\vec  r,\vec  r^{\
\prime}}\vec  \varphi_ \alpha \left(\vec  r \right)\cdot \left(\Delta^
\alpha_{\vec  r\vec  r^{\ \prime}} \right)^{-1}\cdot\vec  \varphi_ \alpha
\left(\vec  r^{\ \prime} \right) \eqno (2.7.b) $$
where the operator $ \Delta^ \alpha_{\vec  r\vec  r^{\ \prime}} $ is 1 if $
\vec  r $ and $ \vec  r^{\ \prime} $ are nearest-neighbour in
direction $ \alpha , $ and 0 otherwise.
Note again that we have omitted the denominator in (2.7.a), since it goes to
unity when $n \to 0$.
\par
In order to prove the equivalence of (2.6) and (2.7), we will use again
Wick's theorem: we define, from (2.7.b), the elementary contraction:
$$ \charlvupup{ \dasharrowfill}{ \varphi^{ (u)}_\alpha \left(\vec  r
\right)\cdot \varphi^{ (v)}_\gamma \left(\vec  r^{\ \prime} \right)} =
\delta_{ uv}\delta_{ \alpha \gamma} \Delta^ \alpha_{\vec  r\vec  r^{\
\prime}} \eqno (2.8) $$
where $ u $ and $ v $ are component indices of $ \vec  \varphi_ \alpha
\left(\vec  r \right), $ running from 1 to $ n. $ \par
Expanding the product over $ \left\{\vec  r \right\} $ in (2.7.a), we must
choose
at each
site $ \vec  r $ either a term of the form $ {1 \over 2} \vec  \varphi^
2_\alpha \left(\vec  r \right) $ or one of the form $ {\rm e}^{-\beta
\varepsilon_h}\vec  \varphi_ \alpha \left(\vec  r \right)\cdot\vec  \varphi_
\gamma \left(\vec  r \right), $
corresponding respectively to a path going straight through $ \vec  r $ in
direction
$ \alpha , $ or to a path making a turn at $ \vec  r, $ from direction $
\alpha $ to direction $ \gamma . $ By
contracting the fields according to (2.8), we construct a sum over all self
avoiding compact closed paths with appropriate weights, with an additional
factor $ n $ (due to the summation over component index $ u) $ for each closed
loop. As above, we extract the single chain contribution
by taking the
limit $ n=0. $ This concludes the proof.
\par
The mean field equations (saddle point on (2.7.a)) read:
$$ \sum^{ }_{\vec  r^{\ \prime}} \left(\Delta^ \alpha_{\vec  r\vec  r^{\
\prime}} \right)^{-1}\vec  \varphi_ \alpha \left(\vec  r^{\ \prime} \right) =
{\vec  \varphi_ \alpha \left(\vec  r \right)+ {\rm e}^{-\beta \varepsilon_h}
\sum^{ }_{ \gamma (\not= \alpha )}\vec  \varphi_ \gamma \left(\vec  r \right)
\over{ 1 \over 2} \sum^{ }_{ \alpha^{ \prime}}\vec  \varphi^ 2_{\alpha^{
\prime}} \left(\vec  r \right)+{1 \over 2} {\rm e}^{-\beta \varepsilon}
\sum^{ }_{ \gamma^{ \prime} \not= \alpha^{ \prime}}\vec  \varphi_{ \alpha^{
\prime}} \left(\vec  r \right)\cdot\vec  \varphi_{ \gamma^{ \prime}}
\left(\vec  r \right)} \eqno (2.9) $$
\par
Assuming periodic boundary conditions,
the isotropic homogenous mean-field solution takes the form  $ \vec  \varphi_
\alpha
\left(\vec  r \right)=\vec  \varphi $ for any $ \alpha =1,...,d $
and $ \vec  r. $ From (2.9), we obtain:
$$ \varphi^{( 1)} = \sqrt{{ 4 \over d}} \eqno (2.10.a) $$
\par
At this mean-field level, the free energy per site reads:
$$ f = -T\ {\rm log} \left({q(\beta) \over {\rm e}} \right) \eqno (2.10.b) $$
with
$$ q(\beta)  = 2 + 2(d-1)\  {\rm e}^{-\beta \varepsilon_h} \eqno (2.10.c) $$
and $ {\rm e}  = 2.71828...\ . $ Note that $ q(\beta) $ plays the role of an
effective
coordination number . \par
The phase transition of the model described by eq.(2.7) has been
studied in detail in reference [6]. The ground state consists of a
bundle--of--sticks like conformation taking the form of
straight paths which make turns on the surface.  This solution, which has $f=0$
in the large $N$ limit,
 is not obtained from the saddle point
equation (2.9), which is not surprising for a structure which is
essentially one dimensional. It must nevertheless be considered, since
it is clearly the lowest energy state of the chain.
Within our approximations, a first order phase transition must then occur at
a temperature $ T_F $ for which $f=0$, yielding an effective coordination
number $ q(\beta) $ equal to $ {\rm e} $. For $ d=3, $
$ T_F = 0.58\ \varepsilon_h . $ Below
this temperature, the free energy remains equal to zero. \par
Physically, there is a competition between the entropy gain of
making turns, and the corresponding energy loss. At high temperature, the
corners are mobile in the bulk, leading to a liquid like structure, whereas
at low temperature, the system is frozen, in stretched walks, with the
corners expelled on the surface. The two regions are separated by a phase
transition at $ T_F. $
\par The average length of a helix is given by
$$ \ell  = {N\varepsilon _h\over U \left(\beta_ F \right)} \eqno (2.11) $$
where $ U(\beta)  = - {\partial \over \partial \beta}  {\rm log} \ Z $ is the
internal energy. Just above the freezing point, in $ d=3, $ the average
secondary structure persistence length is equal to $ \ell_ F = 3.78, $
and is of ${\cal O} (\ell =N^{1/3}) $ in the low temperature phase.
Note that in this very simplified picture $ \ell_ F $ corresponds to a
typical number of amino-acids per
$ \alpha $-helix of the order of 15, since one link corresponds to a
helical turn, that is 3.6 amino acids.
\par Since the density of
corners is extensive (i.e. ${\cal O}(N)$ for $T > T_F$),
 and drops to zero (i.e. ${\cal O}(N^{1-1/d})$)
for $ T < T_F$, the freezing transition is a first order in the
thermodynamic limit.

This model can be easily extended to incorporate the existence of
$\beta$-sheets [7]: the physics of the latter model is very similar to that of
the previous model (2.2), namely, a liquid-like high temperature phase
with no definite $ \beta $-sheet structure, and a low temperature frozen
phase, consisting of stacks of parallel $ \beta $-sheets.  \par \medskip
\vfill\eject
\noindent {\bf 3.\nobreak\ VACANCIES AND GEOMETRICAL FRUSTRATION: A
MEAN-FIELD TREATMENT}
\par
We wish now to explore the full phase diagram of the protein chain,
including its disordered (denaturated) coil state. For the sake of
simplicity, we will generalize the approach of section 2 to the
case of helix--like structures (again, the extension to the $\beta$-sheet case
is straightforward).
We consider a $ d$-dimensional hypercubic lattice of
$\Omega=L^d$ sites, and a polymer chain of $N$ monomers, each monomer
being at a lattice site. Following the preceding discussion,
 we attribute an energy
penalty $\varepsilon_h$ for each corner (i.e., when two consecutive
links are not aligned). Furthermore, we now introduce a finite Van der
Waals attractive energy $\varepsilon_v$ for each pair of non
consecutive monomers separated by one lattice spacing. We emphasize
once more that this model aims at a simplified description of the
solvent conditions.\par
The partition
function reads
\par
\smallskip
$$ {\cal Z }_N= \sum_{(n_{\vec r}=0,1)}\sum^{ }_{\{{\cal
H}_N\}}
 {\rm e}^{- \beta \varepsilon_h N_c\{{\cal
H}_N\}}{\rm e}^{-N\beta \varepsilon_v}
{\rm e}^{{1\over 2}\beta \varepsilon_v\sum_{\vec r \vec r'}n_{\vec
r}\Delta_{\vec r \vec r'}n_{\vec r'}}\ \delta(\sum_{\vec r}n_{\vec r}-N)
 \eqno (3.1) $$
where $n_{\vec r}$ denotes the occupation number of lattice site $\vec
r$, and ${\{\cal H}_N\}$ denotes now a self avoiding path of $N$
monomers on the lattice. The constant factor ${\rm e}^{- N \beta
\varepsilon_v}$ in eq.(3.1) ensures that one does not count the
attractive energy between consecutive monomers. Following
section 2, it is easy to reformulate this problem by introducing at
each site $\vec r$ of the lattice (i) $n$-components fields
 ${\vec  \varphi_ \alpha \left(\vec  r \right)}$, with
$\alpha=1,....,d$ (ii) a scalar field $\psi_{\vec r}$ needed to
perform the Hubbard-Stratanovich transformation on the occupation
numbers $n_{\vec r}$. Introducing a fugacity $K$ per monomer,
the grand canonical partition
function $Z(K)$ may be written:

$$ Z(K) = \dlowlim{ {\rm lim}}{n \longrightarrow 0} {1 \over n} \int^{ }_{ }
\prod^{ }_{\vec  r} {\rm d}\psi_{\vec r}\prod^ d_{\alpha =1} {\rm d}\vec
\varphi_ \alpha
\left(\vec  r \right) {\rm e}^{-A_G} \prod^{ }_{\vec  r} \left[{1 + K
{\rm e}^{-\beta \varepsilon_v}{\rm e}^{\sqrt{\beta
\varepsilon_v} \psi_{\vec r}} B_{\vec r}} \right] \eqno (3.2.a) $$
with
$$ A_G = {1 \over 2}\sum^{ }_{\vec  r,\vec  r^{\
\prime}}\psi_{\vec r}\left(\Delta
_{\vec  r \vec  r'} \right)^{-1}\psi_{\vec r'}+
{1 \over 2} \sum^ d_{\alpha =1} \sum^{ }_{\vec  r,\vec  r^{\
\prime}}\vec  \varphi_ \alpha \left(\vec  r \right)\cdot \left(\Delta^
\alpha_{\vec  r\vec  r'} \right)^{-1}\cdot\vec  \varphi_ \alpha
\left(\vec  r' \right) \eqno (3.2.b) $$
and
$$ B_{\vec r} = \left[{1 \over 2}
\sum^ d_{\alpha =1}\vec  \varphi^ 2_\alpha \left(\vec  r \right)+ {\rm
e}^{-\beta \varepsilon_h} \sum^{ }_{ 1\leq \alpha <\gamma \leq d}\vec  \varphi_
\alpha \left(\vec  r \right)\cdot\vec  \varphi_ \gamma \left(\vec  r \right)
\right] \eqno (3.2.c) $$
it is easy to check, using the approach of section 2, that ${\cal
Z}_N$ in (3.1) is the coefficient of $K^{N}$ in eq.(3.2); we may alternatively
write
$$ N= {\partial \log Z(K)\over \partial \log K}\eqno(3.3)$$
Our strategy is now straightforward: we evaluate (3.2) via an isotropic
 saddle point method and periodic boundary conditions
 ($\vec \varphi_{\alpha}(\vec r)=\vec \varphi;\  \psi_{\vec r}=\psi$). With
these hypothesis, eq.(3.2.a) gives
$$ \log Z(K) = \Omega \left[-{1\over 2}({1\over 2 d}\psi^{2}+{d\over 2}{\vec
\varphi}^2)+\log  ({1 + K
{\rm e}^{-\beta \varepsilon_v}{\rm e}^{\sqrt{\beta
\varepsilon_v} \psi} B} )\right] \eqno (3.4) $$
where $B$ is now space independent and given by (3.2.c). Breaking the $
O(n) $ symmetry as in section 2, the saddle point equations reads
$$ {\psi\over (2d)}  = {(d {\varphi}^2) \over 2}\ {K\sqrt{\beta
\varepsilon_v}\  {\rm e}^{-\beta \varepsilon_v }{\rm e}^{\sqrt{\beta
\varepsilon_v }\psi} \left(1+(d-1)\ {\rm e}^{-\beta
\varepsilon_h}\right) \over D}
\eqno (3.5.a) $$
$$ {d\varphi\over 2}  = (d \varphi)\ {K {\rm e}^{-\beta \varepsilon_v }{\rm
e}^{\sqrt{\beta
\varepsilon_v }\psi} \left(1+(d-1)\ {\rm e}^{-\beta
\varepsilon_h}\right) \over D}
\eqno (3.5.b) $$
with
$$D = 1 + {(d \varphi^2)\over 2}\ {K {\rm e}^{-\beta \varepsilon_v }{\rm
e}^{\sqrt{\beta
\varepsilon_v }\psi} \left(1+(d-1)\ {\rm e}^{-\beta
\varepsilon_h}\right) }
\eqno (3.5.c) $$
Equation (3.3) in turn gives
$$ N  = \Omega\ {(d {\varphi}^2) \over 2}\ {K {\rm e}^{-\beta \varepsilon_v
}{\rm e}^{\sqrt{\beta
\varepsilon_v }\psi} \left(1+(d-1)\ {\rm e}^{-\beta
\varepsilon_h}\right) \over D}
\eqno (3.5.d) $$
Introducing the monomer density $\rho={N\over \Omega}$, we have from
equations (3.5)
$$\varphi^2={4\over d}\  \rho \eqno (3.6.a)$$
$$\psi={2d} {\sqrt{\beta \varepsilon_v}}\ \rho \eqno (3.6.b)$$
$$D={1\over (1-\rho)} \eqno (3.6.c)$$
leading to a free  energy per monomer :
$$ f =- {T\over N}\log {\cal Z }_N\ = - {T\over N}\log
\left({Z(K)\over K^N}\right) \eqno (3.7.a) $$
We finally obtain
$$
f=-T \log \left({q(\beta) \over e}\right) +T {(1-\rho)\over \rho}
\log(1-\rho)
+ \varepsilon_v-d\varepsilon_v \rho
\eqno(3.7.b)
$$
\par
This equation is reminiscent of a Flory-Huggins formula [13] for a
single chain: the first term represents the entropy of the collapsed
chain (see eq.(2.10.b)), the second term accounts for the entropy of
vacancies, and the last two terms represent the attractive energy.
Note that, we work here with the free energy per monomer rather than
with the free energy per site.\par
As discussed in section 2, the fully stretched (or crystalline) phase makes
turns at
the ``surface'', among the vacancies; this phase is not obtained in the saddle
point approximation. Its free energy per monomer is easily obtained as
$$g=- (d-1)\ \varepsilon_v
\eqno (3.8)$$
since $\varepsilon_v$ is the energy between two neighbouring non
consecutive monomers. Note that for density of vacancies $\rightarrow 0$,
i.e. $\rho=1$, the analysis of section
2 is recovered: one may also argue that the limit $\varepsilon_v \to
\infty$
 in equation (3.7.b) implies that the phase with $\rho=1$ corresponds
to the minimum of the free energy $f$. \par
Since our study encompasses the study of the high temperature phase as
well, it is also interesting to consider the value of the free energy $f$
for the limit $\rho=0$, that we call $f_0$. We get
$$
f_0=-T \log q(\beta) +\varepsilon_v,
\eqno(3.9)
$$
that is the free energy of the swollen coil, which corresponds, in our
approximations, to the entropy of a Brownian walk on a lattice with
temperature dependent connectivity.
\par
We now
consider the phase diagram implied by the study of equations (3.7.b)
(3.8), and (3.9).
\par Broadly speaking, one should compare the free energies per monomer $f$
and $g$. A first order freezing of the type described previously will
occur when $g \le f$. We have chosen to present our result in the
$(t,x)$ plane, with $t={T \over \varepsilon_v}$ and $x={\varepsilon_h \over
\varepsilon_v}$. Beside
the extremal points $\rho=0$ and $\rho=1$, we have to consider the
solution $\rho=\rho^{\star}$ with
$$ \left({{\partial} f\over \partial \rho} \right)_{\rho=\rho^{\star}}=0
\eqno (3.10.a)$$
yielding
$$ {\log \ (1-\rho^{\star}) \over \rho^{\star}}+1+ \rho^{\star}{d \over t}=0
\eqno(3.10.b)$$
\vskip 2mm
\item (a) the $\theta$ collapse transition\par
{}From eq.(3.10.b), one may
get a non-zero density $\rho^{\star}$, below the $x$-independent
$\theta$ temperature $t_{\theta}=2d$, i.e. $t_{\theta}=6$ for $d=3$.
\item (b) the freezing transition\par
One has to compare
$f(\rho^{\star})$ and $g$, where $\rho^{\star}$ is given by eq.(3.10.b). It
is easy to see that for $x \to \infty$, we have at the transition
$\rho_{\infty}^{\star}=0.5$, and $t_{\infty}={d \over (4 \log2-2)}$,
yielding $t_{\infty} \simeq 3.88$ for $d=3$.
For $x \to 0$, we recover the result of section 2,
namely $\rho_0^{\star}=1$ and $t_0 ={x / {\log \left({2(d-1) \over
(e-2)} \right)}}$, i.e. $t_0 \simeq 0.58x$ for $d=3$.
\par
\vskip 2mm
Note that in the mean-field treatment of our
model, it is not possible to obtain a direct
transition from the high temperature phase to the crystalline ground
state. This transition would occur for $f_0=g$, and the resulting
temperature is lower than $t_{\theta}$, as seen from eqs(3.8) and
(3.9). The fact that one has to go through a dense intermediate phase
can be traced back to the vacancy entropy in eq (3.7.b). The resulting
phase diagram is shown in Figure 1. We now study corrections to the mean-field
picture.
\medskip

\noindent {\bf 4.\nobreak\ VARIATIONAL BOUNDS AND MONTE CARLO
SIMULATIONS: THE MODIFIED PHASE DIAGRAM}
\par
The $\theta$ collapse transition is a 2nd order transition, and is
thus not very accurately described by mean-field methods, whereas the
freezing transition being first-order may be quite well described
within this framework. For example, we find too high a value of the $x$
-independent collapse temperature ($t_{\theta}=6$ for $d=3$).
Numerical estimates [19] give $t_{\theta} \simeq 3.64$ for $x=0$,
which may imply a modification of the phase diagram of figure 1.

A variational bound to the free energy may be easily obtained. Indeed,
from equation (3.2.c) we have (using Schwarz inequality):
$$
\eqalign{
B_{\vec r}
&= {d \over 2} (1 - {\rm e}^{-\beta \varepsilon_h})
\ \left ({1 \over d} \sum^ d_{\alpha =1}\vec  \varphi^ 2_\alpha
\left(\vec  r \right) \right )+ {{\rm
e}^{-\beta \varepsilon_h} \over 2} (\sum^{ }_{ 1\leq \alpha \leq d}\vec
\varphi_
\alpha \left(\vec  r \right))^2  \cr
&\ge { q(\beta) \over 2d} \ \left( {1\over 2}(\sum^{ }_{ 1\leq \alpha \leq
d}\vec  \varphi_
\alpha \left(\vec  r \right))^2  \right )\cr
}
\eqno (4.1.a)
$$
where $q(\beta)$ is the effective connectivity of the lattice, as
given by (2.10.c).
{}From (4.1.a) we get the bound
$$
Z \ge \left( {q(\beta) \over 2d} \right )^N Z_{\theta}
(\varepsilon_v,T)
\eqno(4.1.b)
$$
where $Z_{\theta} (\varepsilon_v,T)$ is the partition function of a
chain with no curvature energy, with collapse energy $\varepsilon_v$.
We first note on this equation that the r.h.s. displays a collapse
transition at the $\theta$ temperature of the ``pure'' system, that is the
system with no stiffness. This result is consistent with the result of
section 3.\par
As mentioned above, the ``pure'' $\theta$ point has been calculated
numerically in three dimensions, leading to a temperature much lower than the
mean-field estimate ($t_{\theta} \simeq 3.64$ vs. a mean-field estimate
of $t_\theta = 6$). This effect is partly due
to the fact that the $\theta$ point is a second order phase
transition, which is dominated by fluctuations, which are not well
taken into account at the mean-field level.

On the other hand, the freezing transition is a first order
transition, and thus is not driven by large fluctuations. The
mean-field approximation is then much more reliable.

If we take these two facts seriously, we obtain the phase diagram
represented on figure 2. One qualitative modification of the phase
diagram is the possibility of a direct freezing transition, without
passing through a $\theta$ collapse. Indeed, defining the critical value $x_c$
as the intersection between the line of the $\theta$ collapse
($t=3.64$) and the freezing curve, we see that for:
\item {$x < x_c$:} the transition to the native (crystalline) state
proceeds through a $\theta$ collapse, followed by a freezing
transition. There is a whole region in temperature, where the system
is collapsed, but with no definite secondary structure, and this may be
identified with a ``molten globule'' state.\hfill
\item {$x > x_c$:} the freezing transition occurs at a higher
temperature than the $\theta$ collapse, and thus the system will
directly freeze, from the swollen coil to the native state. \hfill
\par
\noindent
Using equations (3.7.b), (3.8), and (3.10) together with the three
dimensional numerical value of $t_{\theta}$, we find $x_c \simeq 15$.\par
\noindent
\vskip 4mm
We have checked these predictions numerically by performing Monte
Carlo simulations. We have used a Monte Carlo growth method [20],
which has proven both fast and reliable for calculations of
thermodynamical properties of random chains. We grow populations of
$M=20000$ chains of length $N=200$, for various values of
$t=T/\varepsilon_v$ and $x=\varepsilon_h/\varepsilon_v$. There is of
course a degradation of the statistics in the frozen phase, and we
have thus restricted the parameter $x$ to 3. The numerical results are
plotted on figure 2. Each calculated point is represented by a star. We see
that the agreement is quite good, and seems to confirm the theoretical
phase diagram predicted above.

\vskip 3mm
\noindent {\bf 5.\nobreak\ CONCLUSIONS}
\par
We have studied a simplified homopolymer model
to describe the interplay  between compactness and local order in a
protein chain. In  a mean-field treatment of this model,
the native state can be reached only after a $\theta$ collapse
transition towards a liquid globule (the collapse transition
temperature being independent of the stiffness (or curvature) energy).
In this globule, helix--like secondary structures progressively
grow and ultimately freeze. These results are not capable of representing
the traditional two-state coil $\leftrightarrow$  native renaturation
transition for globular proteins. \par
 We have argued that this disagreement may be partially explained by the
mean-field treatment. Using more reliable estimates of the  three
dimensional $t_{\theta}$ temperature [19], we have used variational
bounds, together with Monte Carlo calculations, to show that (i)
$t_{\theta}$  is indeed independent of $\varepsilon_h$  (ii) the phase
diagram may be qualitatively modified  for large enough values of
$x=\varepsilon_h/\varepsilon_v$,
allowing for a unique transition from the coil towards the native
state. In our model, we get a very high value for the critical $x_c$,
separating the two transitions regime from the unique ``coil-native state
renaturation transition''($x_c \simeq 15$). This value, which shows
that a large portion of the phase diagram is occupied by a compact
liquid or molten globule phase, is certainly
not realistic for globular proteins.
\vskip 2mm
 \par
Finally, we think that dealing with an homopolymer model may be inappropriate
to describe the renaturation transition of globular proteins.  The latter has
been modeled in a simplified theoretical approach by Finkelstein and
Shakhnovich [14,15] and by Alonso, Dill and Stigter [16].
Transitions of this type have also been studied analytically in a
random heteropolymer model [17]. Such two-state
transitions have their origin in the heteropolymer nature of proteins for
which hydrophobic residues become segregated within the core of the
protein, while hydrophilic residues are in contact with solvent.
\par
Our results show that the homopolymer model has a molten globule like
thermodynamic phase.  Although the model has no side chains and is hence
incapable of including the effects of side chain entropy on the
native $\leftrightarrow$
molten globule transition, it does reveal another entropic contribution
which has not been considered in this context by previous workers:
the entropy of backbone disordering.

Thus, even though side chain rotational entropy is undoubtedly important
in the molten globule state, we have shown that backbone disordering may
be capable of driving a first order transition, and hence that side chain
entropy need not necessarily be the driving force for the transition.  Thus
backbone disordering provides a complementary mechanism to the discussion
of Bromberg and Dill [18], who argue that the onset of side chain disordering
may not
be sufficiently sharp to drive the native $\leftrightarrow$
 molten globule transition.

Last but not least, it is clear that our results pertain to the case of
semi- flexible polymers.  The intermediate compact phase is then likely
to posess orientational (e.g. nematic) order [9,10].
\vfill\eject
\centerline{{\bf FIGURE CAPTIONS}}
\noindent {Figure 1\nobreak\ :} Mean field phase diagram in the ($t,x$)
plane. The solid line denotes the continuous $\theta$ transition from
the coil (C) to the liquid globule (G). The
dashed line denotes the first order freezing transition to the native
state (N).\par
\vskip 2mm
\noindent {Figure 2\nobreak\ :}   Numerical three dimensional phase
 diagram in the ($t,x$)
plane. The solid line denotes the continuous $\theta$ transition from
the coil (C) to the liquid globule (G). The
dashed line denotes the first order freezing transition to the native
state (N).  The stars show the results of the Monte Carlo
calculations. The multicritical point corresponds to ($t_c \simeq 3.6, $ $x_c
\simeq 15$).
\vfill\eject

\centerline{{\bf REFERENCES}}
\medskip
\noindent [1] T.E.Creighton (editor), \lq\lq
Protein Folding\rq\rq , W.H. Freeman, New York, (1992)\par

\noindent [2] O.B.Ptitsyn, in reference [1], p.243\par

\noindent [3] Fink, A L, {\sl Ann. Rev. Biophys. and Biomol. Struct.},
{\bf 24}, 495-522, (1995)\par

\noindent [4] P.J.Flory , {\sl Proc. Roy. Soc.\/}
{\bf A234}, 60, 73, (1956)\par

\noindent [5] Kolinski A, Skolnick J and Yaris R.  {\sl Proc.
Natl. Acad. Sci. (USA)} {\bf 83}, 7267-71, (1986)\par

\noindent [6] J.Bascle, T.Garel and H.Orland,  {\sl J.Phys.A} {\bf 25},
L1323, (1992)\par

\noindent [7] J.Bascle, T.Garel and H.Orland, {\sl J.Phys.(France)} {\bf
II}, 3,
245, (1993)\par

\noindent [8] S. Doniach, J. Bascle, T. Garel and H. Orland, submitted
for publication (1995)\par

\noindent [9] H.Saleur, {\sl J.Phys.A} {\bf 19}, 2409, (1986)\par
\noindent [10] M.Dijkstra and D.Frenkel,
{\sl Phys.Rev.\/} {\bf E50}, 349, (1994)\par
\noindent [11] J.des Cloizeaux and G.Jannink, \lq\lq
Les Polym\`eres en
Solution\rq\rq\ Eds. de Physique, Les Ulis, (1987)\par

\noindent [12] H.Orland, C.Itzykson and C.De
Dominicis, {\sl J. Phys.(France)\/} {\bf 46},
L353, (1985)\par

\noindent [13] P.G.de Gennes, \lq\lq
Scaling concepts in polymer physics\rq\rq , Cornell University Press,
Ithaca, (1979)
\par

\noindent [14] A.V.Finkelstein and E.I.Shakhnovich, {\sl Biopolymers\/}
{\bf 28}, 1682, (1989)\par

\noindent [15] E.I.Shakhnovich and A.V.Finkelstein, {\sl Biopolymers\/}
{\bf 28}, 1667, (1989)\par

\noindent [16]D.O.V.Alonso, K.A.Dill and D.Stigter
{\sl Biopolymers}  {\bf 31}:1631-1650, (1991).\par

\noindent [17] T.Garel, L.Leibler and H.Orland, {\sl J.
Phys.(France)\/}
 {\bf II}, 4, 2139, (1994)\par

\noindent [18] S.Bromberg and K.A.Dill,
{\sl Protein Science} {\bf 3}, 997-1009, (1994)\par

\noindent [19] H.Meirovitch and H.A.Lim,
{\sl J.Chem.Phys.} {\bf 92}, 5144-54, (1990)\par

\noindent [20] T.Garel and H. Orland,
{\sl J.Phys.A} {\bf 23}, L621-24, (1990)\par

\vfill\eject

\end